\documentclass[review]{elsarticle}
\usepackage{graphicx}
 \usepackage[usenames]{color}
\usepackage{lineno,hyperref}
\usepackage{lineno}
\usepackage{geometry}
\journal{JINST}

\begin{document}

\begin{frontmatter}
\title{{\bf ML-based muon identification using a FNAL-NICADD scintillator chamber for the MID subsystem of ALICE\,3}}

\author[icn_unam]{Jes\'us~Eduardo~Mu\~noz~M\'endez}
\author[icn_unam]{Antonio~Ortiz}
\author[icn_unam]{Antonio~Paz}
\author[icn_unam]{Paola~Vargas~Torres}
\address[icn_unam]{Instituto de Ciencias Nucleares, Universidad Nacional Aut\'onoma de M\'exico, Mexico}

\author[if_unam]{Ruben~Alfaro~Molina}
\author[if_unam]{Laura~Helena~Gonz\'alez~Trueba}
\author[if_unam]{Varlen~Grabski}
\address[if_unam]{Instituto de F\'isica, Universidad Nacional Aut\'onoma de M\'exico, Mexico}

\author[fcfm_buap]{Arturo~Fern\'andez~T\'ellez}
\author[fcfm_buap]{Hector~David~Regules~Medel}
\author[fcfm_buap]{Mario~Rodr\'iguez~Cahuantzi}
\author[fcfm_buap]{Guillermo~Tejeda~Mu\~noz}
\author[fcfm_buap]{Yael~Antonio~V\'asquez~Beltran}
\address[fcfm_buap]{Facultad de Ciencias F\'isico Matem\'aticas, Benem\'erita Universidad Aut\'onoma de Puebla, Mexico}

\author[itn]{Juan~Carlos~Cabanillas~Noris}
\address[itn]{Instituto Tecnol\'ogico de Culiac\'an, Tecnol\'ogico Nacional de M\'exico}

\author[prague]{Solangel~Rojas~Torres} 
\address[prague]{Czech Technical University in Prague, Prague, Czech Republic}

\author[wigner]{Gergely~G\'abor~Barnaf\"oldi}
\author[wigner]{D\'aniel~Sz\'araz}
\author[wigner]{Dezs{\H o}~Varga}
\author[wigner]{R\'obert~V\'ertesi}

\address[wigner]{HUN-REN Wigner Research Centre for Physics, Hungary}

\author[chicago]{Edmundo~Garc\'ia~Solis} 
\address[chicago]{Chicago State University, Chicago, USA}


\cortext[mycorrespondingauthor]{antonio.ortiz@nucleares.unam.mx}


\begin{abstract}

The ALICE Collaboration is planning to construct a new detector (ALICE\,3) aiming at exploiting the potential of the high-luminosity Large Hadron Collider (LHC). The new detector will allow ALICE to participate in LHC Run 5 scheduled from 2036 to 2041. The muon-identifier subsystem (MID) is part of the ALICE\,3 reference detector layout. The MID will consist of a standard magnetic iron absorber ($\approx4$ nuclear interaction lengths) followed by muon chambers. The baseline option for the MID chambers considers plastic scintillation bars equipped with wave-length shifting fibers and readout with silicon photomultipliers. This paper reports on the performance of a MID chamber prototype using 3\,GeV/$c$ pion- and muon-enriched beams delivered by the CERN Proton Synchrotron (PS). The prototype was built using extruded plastic scintillator produced by FNAL-NICADD (Fermi
National Accelerator Laboratory  - Northern Illinois Center for Accelerator and Detector
Development). The prototype was experimentally evaluated using varying absorber thicknesses (60, 70, 80, 90, and 100\,cm) to assess its performance. The analysis was performed using Machine Learning techniques and the performance was validated with GEANT 4 simulations. Potential improvements in both hardware and data analysis are discussed. 

\end{abstract}

\begin{keyword}
\texttt{Upgrade,} \texttt{Scintillators,} \texttt{SiPM,} \texttt{Machine Learning,} \texttt{ALICE experiment} 
\end{keyword}

\end{frontmatter}

\section{Introduction
\label{sec:Introduction}}
The ALICE apparatus has been successful in studying  the properties of the strongly-interacting quark--gluon plasma (sQGP), the deconfined state of strongly-interacting matter~\cite{ALICE:2022wpn,Bala:2016hlf}. However, some fundamental questions on the sQGP and other aspects of the strong interaction will remain open after the LHC Runs 3 and 4 (2022-2033)~\cite{alice3loi}. To address these questions and fully exploit the potential of the heavy-ion collisions at the LHC during Run 5 (2036-2041), a completely new apparatus named ALICE\,3 is proposed~\cite{Dainese:2925455}. ALICE\,3 consists of a silicon-pixel tracking system with unique pointing resolution (10\,$\mu$m at $p_{\rm T}=200$\,MeV/$c$) over a large pseudorapidity range ($-4 < \eta < 4$), complemented by subsystems for particle identification, including silicon time-of-flight layers, a ring-imaging Cherenkov detector, a muon-identification detector (MID),  and a forward photon conversion tracker. A new superconducting solenoid magnet with a field strength of 2\,T will provide a transverse momentum resolution similar to that of the present ALICE detector in the central region, as well as an excellent momentum resolution at forward rapidity.

The MID subsystem will provide muon tagging for particles reconstructed in the tracker. Its goal is the identification of $J/\psi$ down to $p_{\rm T} = 0$ in the rapidity range $|y|<1.24$ through the reconstruction of its dimuon decay channel. A competitive detection efficiency requires identifying muons down to momenta of $p \approx 1.5$~GeV$/c$ at $\eta \approx 0$. Simulations of the hadronic background in pp collisions at the highest LHC energy, as well as in 0–10\% Pb--Pb collisions at $\sqrt{s_{\rm NN}} = 5.5$~TeV were performed~\cite{alice3loi,Dainese:2925455}. Based on simulations, a standard magnetic steel absorber with a thickness varying from 70~cm at $|z|<100$~cm to 38~cm at $400<|z|<500$~cm (uniform nuclear absorption length of $\approx 4$ along $\eta$) followed by muon chambers with a granularity of $\approx 5\times 5$\,cm$^2$ is sufficient for efficient muon identification from low (1.5~GeV$/c$) to intermediate (8~GeV$/c$) $p_{\rm T}$. 

The baseline option for the MID chamber consists of two layers of plastic scintillator bars (5\,cm wide, 100\,cm length, and 1\,cm thick) equipped with wavelength-shifting (WLS) fibers which are read out with silicon photomultipliers (SiPM). The two layers are
placed perpendicular with respect to each other and separated for an inter-layer gap of 10 cm. Studies using atmospheric muons, as well as in the CERN T10 test-beam facility, have shown that extruded plastic scintillators from FNAL-NICADD~\cite{Pla-Dalmau:2000puk} can provide uniform efficiency ($\approx100\%$) along the 1~m length~\cite{Alfaro:2024sxc}. The typical time resolution of the channel is mostly determined by the length of the scintillator bar, $\approx5/\sqrt{12}$\,ns (a refractive index of 1.5 is assumed within the 1\,m-length bar), and depending on the number of photoelectrons (p.e.), 1.6--2.0~ns can be achieved.  This technology meets the required timing performance, high efficiency, and uniform detector response~\cite{Alfaro:2024sxc}. Regarding radiation tolerance, as discussed in this Ref.~\cite{Garutti:2018hfu}, the main issue with irradiated SiPM is the increase of dark-count rate (DCR) at a relatively low neutron fluences $\Phi_{\rm eq}\approx 10^{11}/{\rm cm}^{2}$, which affects the single p.e. separation from noise. However, recent test beams showed that charged particles with $p_{\rm T}>0.5$~GeV$/c$ produce more than 40\,p.e. in the scintillator paddles used in the present study~\cite{Alfaro:2024sxc}. Therefore, the DCR increase would not reduce the muon detection efficiency, since the signals are much larger than that of a few photoelectrons.  Concerning the total ionizing dose (TID), FNAL-NICADD scintillators have a decrease in light yield of approximately 5\% for a dose of 1~Mrad. However, from FLUKA simulations~\cite{Ahdida:2806210} the total expected TID in Run 5 is significantly below that value. 

This paper reports on test-beam results of a two-layer plastic scintillator chamber. The goal of this work is to present an analysis based on Machine Learning (ML) for muon tagging, as well as the study of the hadron suppression as a function of the absorber thickness. Monte Carlo simulations are compared with measurements using 3\,GeV/$c$ muon- and pion-enriched beams from the CERN PS. Both beams have an unknown percentage of hadrons in the muon-enriched beam and muons in the hadron-enriched beam; this feature of the beam is addressed and investigated with the help of simulations.

The paper is organized as follows, Sec.~\ref{sec:DetectorPrototype} discusses the prototype construction, and Sec.~\ref{sec:ExperimentalSetup} describes the experimental setup at T10. Section~\ref{sec:Methodology} discusses the methodology, namely the GEANT\,4 simulations as well as data analysis using Machine Learning. The results and a summary of the main findings are presented in Sec.~\ref{sec:Results} and Sec.~\ref{sec:Conclusions}, respectively.

\section{Detector prototype construction
\label{sec:DetectorPrototype}}

For this test, a MID chamber prototype of $21.4 \times 21.4$~cm$^2$ geometric area was constructed using FNAL-NICADD scintillators equipped with wavelength-shifting fibers and SiPM for read out. The MID chamber consists of two layers separated by a distance of 10\,cm. Each layer is made with 5 scintillator paddles with a size of $25 \times 4 \times 1$~cm$^3$, the bars in layer 1 are oriented orthogonally to the bars in the second layer. The spacing between adjacent bars is 3.5\,mm.  The scintillator paddles were assembled using the Kuraray WLS fiber model Y-11(200), mounted inside the $2$\,mm groove on the large surface of the scintillator coupled with optical grease, and readout with Hamamatsu S13360-3050CS SiPM coupled on one side of the fiber using optical grease.  The WLS fiber is used to increase the efficiency and homogeneity of detection, thanks to its longer attenuation length of approximately four meters compared to the few centimeters of the FNAL-NICADD scintillator. The SiPM used has a photosensitive area of $3.0\times3.0$\,mm$^{2}$ and pixel pitch of 50\,$\mu$m. Its detection spectral response is located in the wavelength interval 270-900\,nm, with a photon detection efficiency (PDE) of 40\% at peak sensitivity wavelength (450\,nm). The SiPM were connected to the CAEN DT5202 module, a 64-channel front-end unit that provides SiPM bias voltage and readout, thanks to two Weeroc Citiroc-1A ASICs. The two layers are assembled by using a 3D-printed mechanical support that allows to install the electronic-SiPM boards independently to each layer frame. To avoid any light contamination, an aluminum enclosure is used to house both layers, which are kept in place using bolts.

\section{Experimental setup
\label{sec:ExperimentalSetup}}

\begin{figure}[h!]
\centering
\includegraphics[width=0.8\linewidth]{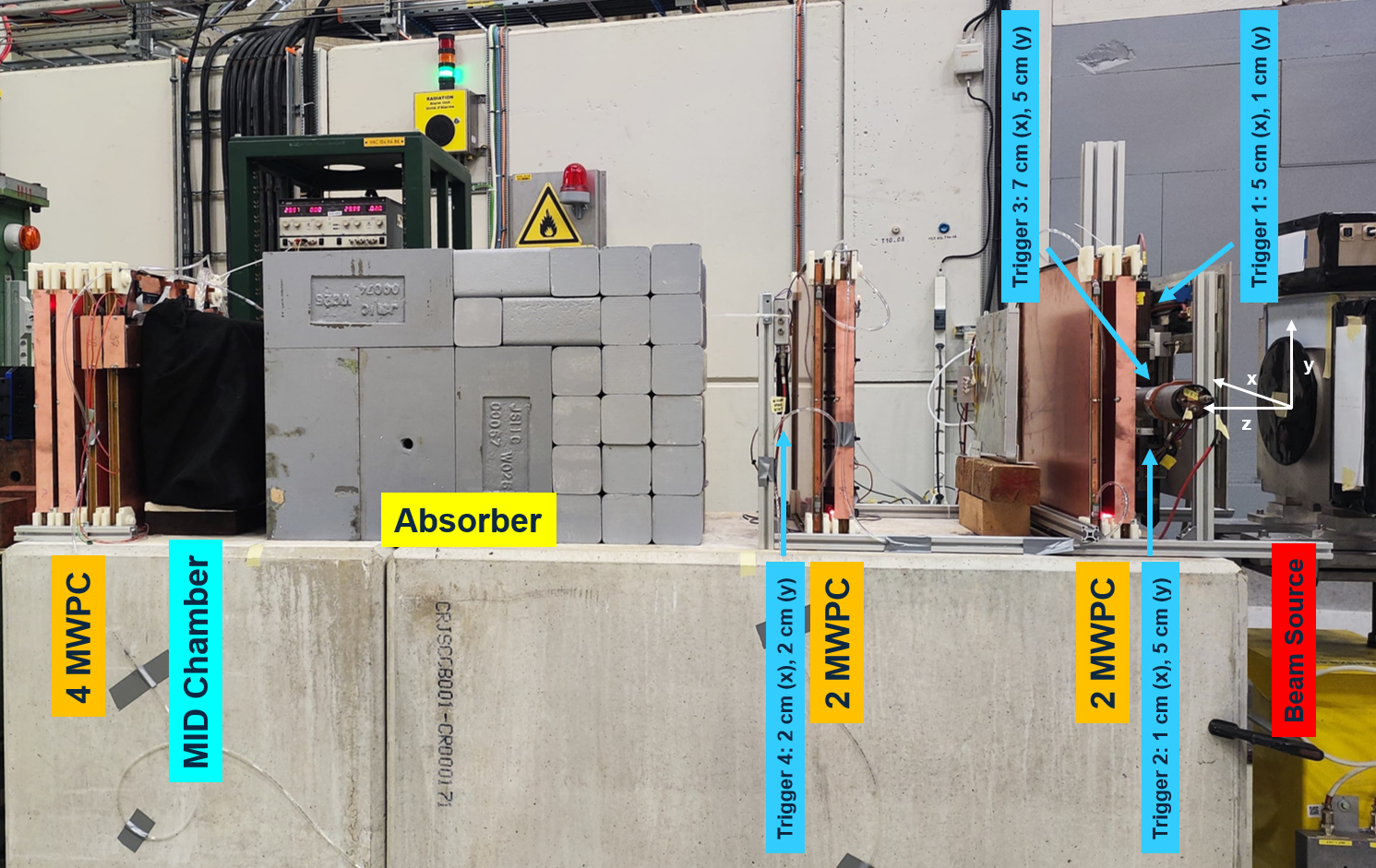}
\caption{Experimental setup. The trigger scintillators are labeled as Tigger~1, Trigger~2, Trigger~3, and Trigger~4. The beam travels from right to left along the $z$ coordinate that is perpendicular to the array of detectors tested at T10.}
\label{fig:1}
\end{figure}

The MID chamber was tested with 3\,GeV/$c$ pion- and muon-enriched beams at the CERN T10 beamline. Figure~\ref{fig:1} shows the setup installed in the beam area.  The setup consisted of four scintillator pallets to provide a clean trigger, an iron absorber of $60\times60$\,cm$^{2}$ transverse area of variable length (from 60 to 100\,cm) followed by the MID chamber. The coordinate system is defined as follows: the beam particles travel along the $z$ axis, the horizontal axis corresponds to the $x$ axis, whereas the vertical axis corresponds to the $y$ axis. 

\begin{figure}[h!]
\centering
\includegraphics[width=0.48\linewidth]{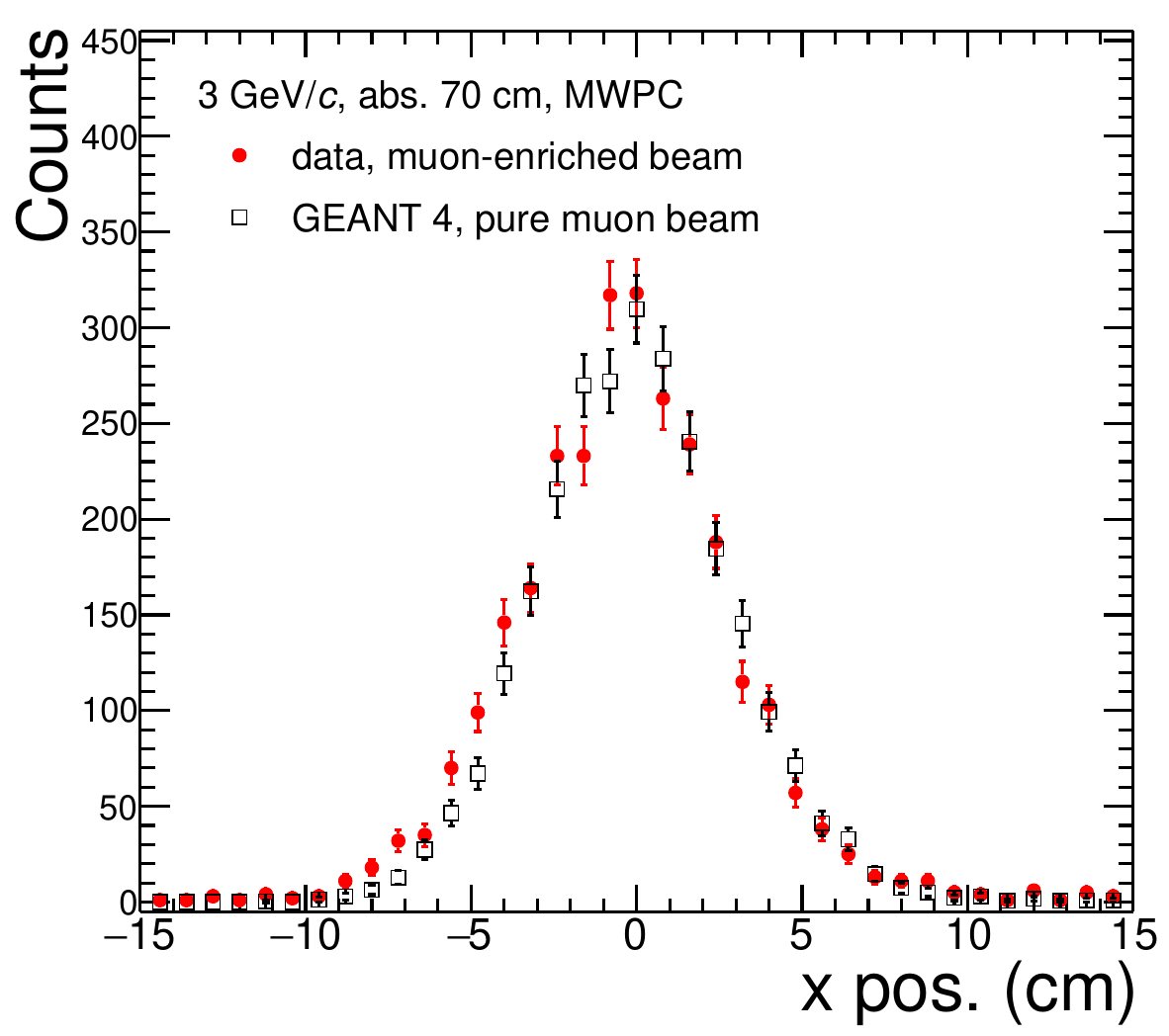}
\includegraphics[width=0.48\linewidth]{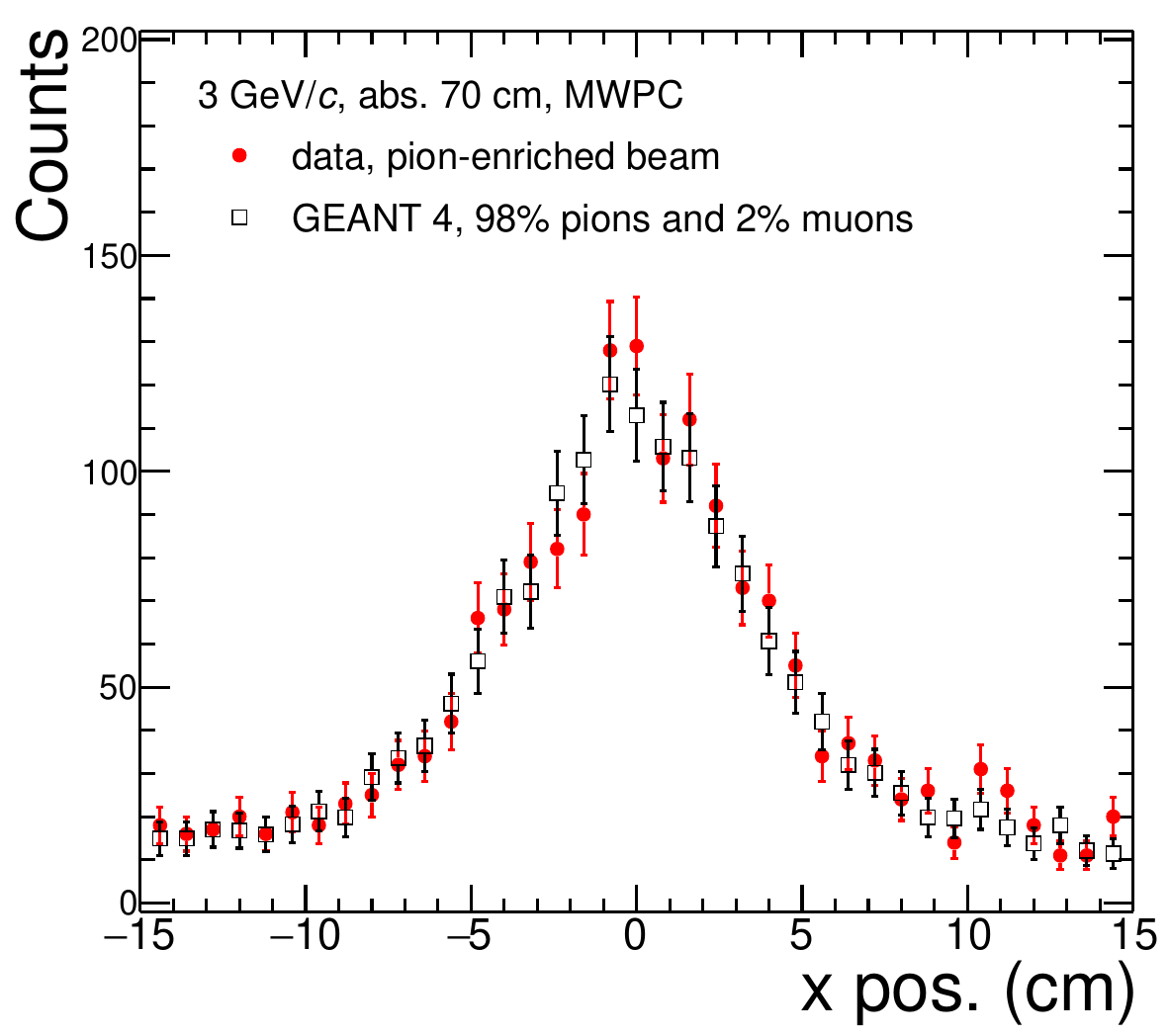}
\caption{The beam profiles for 3\,GeV/$c$ muon- and pion-enriched beams are displayed in the left- and right-hand side plots, respectively. An iron absorber 70\,cm thick is considered. GEANT\,4 simulations assumed a pure muon beam, whereas the pion beam considers a 2\% muon contamination. See the text for details behind this assumption.}
\label{fig:1c}
\end{figure}

The trigger signal was provided by coincidence among four scintillator paddles (hereinafter called trigger scintillators), which were located in front of the absorber. They are indicated as Trigger 1, Trigger 2, Trigger 3, and Trigger 4; and have an active transverse area of $5\times1$\,cm$^{2}$, $5\times1$\,cm$^{2}$, $5\times7$\,cm$^{2}$ and $2\times4$\,cm$^{2}$, respectively. The position of the paddles with respect to the beam is indicated in Figure~\ref{fig:1}. Triggers 1 and 2 were installed close to the beam source, oriented orthogonally to each other to provide a 1~cm$^2$ narrow area trigger.  The second set of triggers was installed at 21.5\,cm (Trigger 3) and 87.7\,cm (Trigger 4) just before the particles enter the absorber.  The trigger scintillators are coupled to Hamamatsu R3478 photomultipliers with short light guide, resulting in high efficiency and a high signal-to-noise ratio with a time resolution of approximately 1\,ns. The trigger signals were processed with a NIM leading edge discriminator (CAEN N840), the output signals were further processed with a coincidence module (LeCroy 622) to define the four-fold coincidence that defined the trigger signal.  The beam profile was measured using an array of eight multiwire proportional chambers (MWPCs)~\cite{VARGA201311}; four were placed in front of the absorber  and the other four were placed behind the absorber. The MID chamber was placed behind the absorber and in front of the second set of four MWPCs. The MID chamber is centered at approximately $x=0.3$\,cm and $y=1.3$\,cm with respect to the beam axis.  The horizontal distributions of the impacts of the beam particles in the detector volume (hits) measured with the MWPCs behind the absorber for 3\,GeV/$c$ particles (muons and pions) are shown in Fig.~\ref{fig:1c}. The shape of the beam profile is well described by GEANT\,4 (version 11.3.0~\cite{GEANT4:2002zbu}) simulation. For pion simulations, a contamination of 2\% muons was assumed. The details behind this assumption will be explained later. It should be noted that the beam is fully contained in the detection area of the MID chamber ($21.4 \times 21.4$~cm$^2$).

\section{Monte Carlo simulations and data analysis
\label{sec:Methodology}}

The simulation of the detector geometry and the propagation of pions and muons through the materials was performed using GEANT\,4. Point-like muon and pion beams with momentum of 3\,GeV/$c$ were considered. The simulation followed the specifications of the prototype: bar spacing of 3.5\,mm, 5 bars in the first layer, and 5 bars in the second layer. The distance between the two layers was 10\,cm, and the distance between the absorber and the first layer was 21\,cm. The bars of the first layer were aligned along the $x$ axis, while the bars of the second layer were aligned along the $y$ axis, so each layer provides the position in $x$ and $y$, respectively. 

\begin{figure}[h!]
\centering
\includegraphics[width=0.48\linewidth]{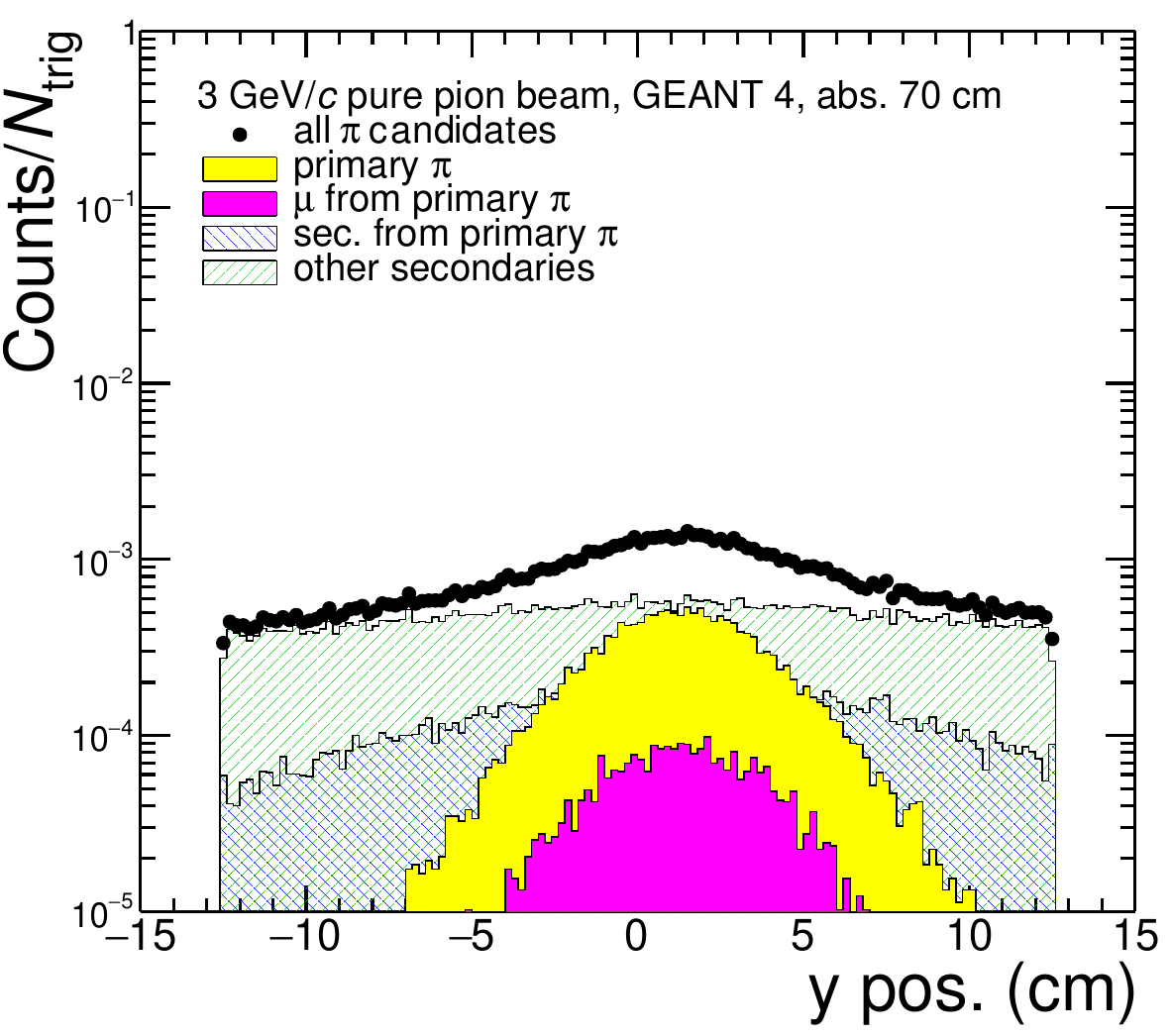}
\includegraphics[width=0.48\linewidth]{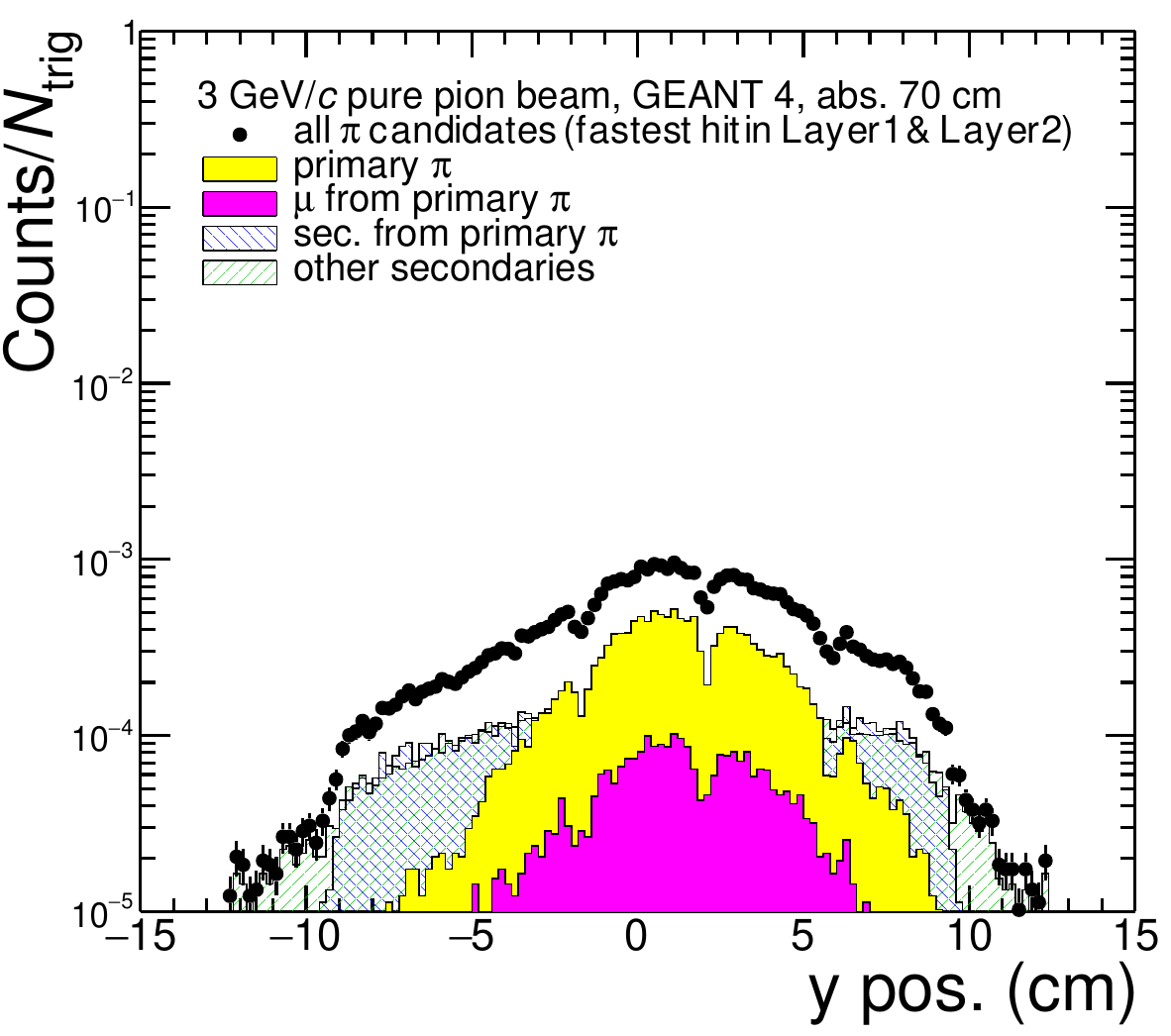}
\caption{(Left) Hit position distribution in the first layer of the MID chamber simulated with GEANT\,4, particles are generated from 3\,GeV/$c$ pions interacting with a 70\,cm thick absorber. The distributions for primary and secondary particles are indicated. (Right) The hit position  distribution for the fastest hit in the first layer, a signal in each layer of the chamber is required as well as a cut in the ToA difference between layers 1 and 2.}
\label{fig:1b}
\end{figure}

Neither the WLS fibers nor the SiPM were explicitly included in the simulations; however, their effects were parametrized. In particular, 8\% of all photoelectrons were considered to mimic the trapping efficiency and reflection in the mirror side of the fiber. The known SiPM PDE was applied to the remaining photons. Since in the experiment time-over-threshold (ToT) was measured instead of charge, ToT was assumed to increase linearly with the p.e. multiplicity ($N_{\rm p.e.}$) up to $N_{\rm p.e.}=72$, beyond its slope is reduced. Regarding the time-of-arrival (ToA), a Gaussian function was fitted to the measured ToA distribution in order to obtain the mean and width of the distributions. The time smearing was introduced using such a parametrization. The hits considered in the analysis are those that deposited an energy greater than 0.15\,MeV (muon efficiency $>99$\% according with GEANT\,4 simulations). Figure~\ref{fig:1b} shows the position distribution of all hits registered in the first layer of the chamber for a 3\,GeV/$c$ pion beam, and considering a 70\,cm thick absorber. Four different contributions are displayed: hits from primary pions, hits produced by muons from primary pions, hits from secondaries (mostly hadrons) from pion interactions with the absorber, and other hadrons. For the latter, the neutron (from thermal to fast-neutron energies) contribution is negligible because their interaction with the scintillator is characterized by low energy deposition as well as signals that arrive at late times ($>10$\,ns). Therefore, they are removed due to the cut on energy deposition. To further reduce the secondary-particle contribution, the hits with the shortest ToA values were selected in layers 1 and 2 of the chamber. In addition,  a cut in ToA was implemented, that is, the time difference between ToA in layers 1 and 2 is kept smaller than 3\,ns, which is around 3 sigma of the measured $({\rm ToA}_{1}-{\rm ToA}_{2})$ distribution. The coincidence between the two layers yields a hit position distribution with a significantly smaller secondary-particle contribution (see the right-hand side of Fig.~\ref{fig:1b}). The distributions exhibit deeps at the $y$ positions where there is no detector in the second layer. For the analysis discussed in the following, this clean sample of hits is used.

\begin{figure}[ht]
\centering
\includegraphics[width=0.90\linewidth]{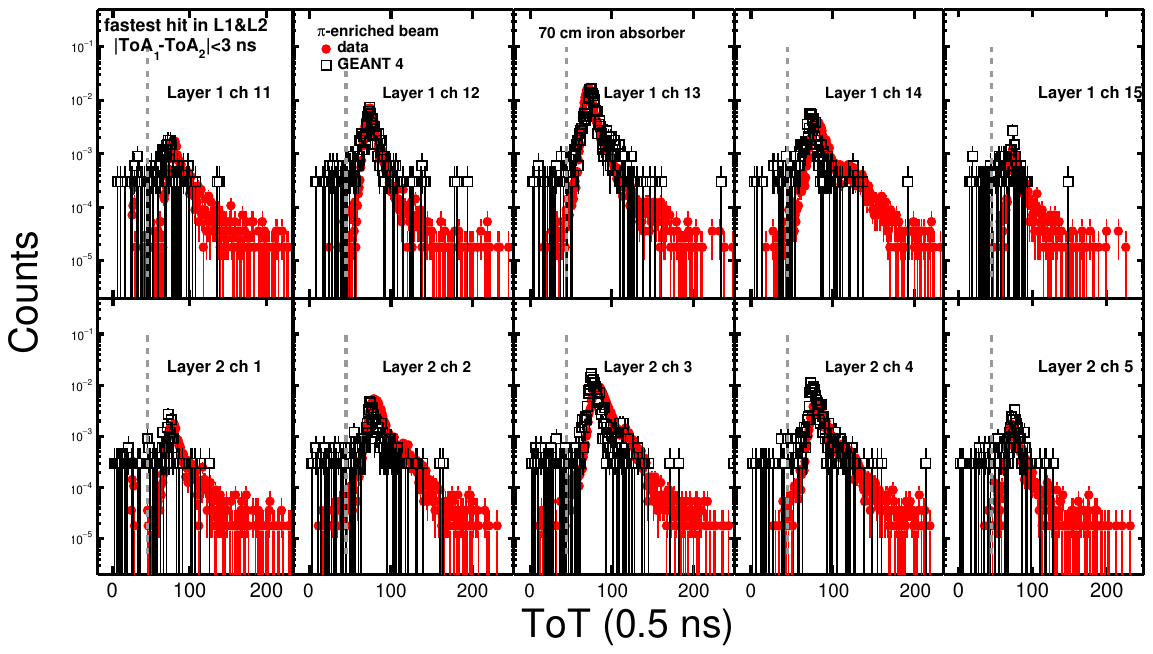}
\caption{Time-over-threshold distributions measured in the 10 channels of the prototype. The upper panel show the results for the first layer, whereas, the bottom panel shows the results for the second layer. The GEANT\,4 results (empty markers) are compared with data (full markers). The measurements were performed using a pion-enriched beam with a momentum of 3\,GeV/$c$. The vertical axis indicates the number of hits normalized to the number of events, which satisfies the selection criteria.}
\label{fig:2}
\end{figure}

\begin{figure}[h!]
\centering
\includegraphics[width=0.90\linewidth]{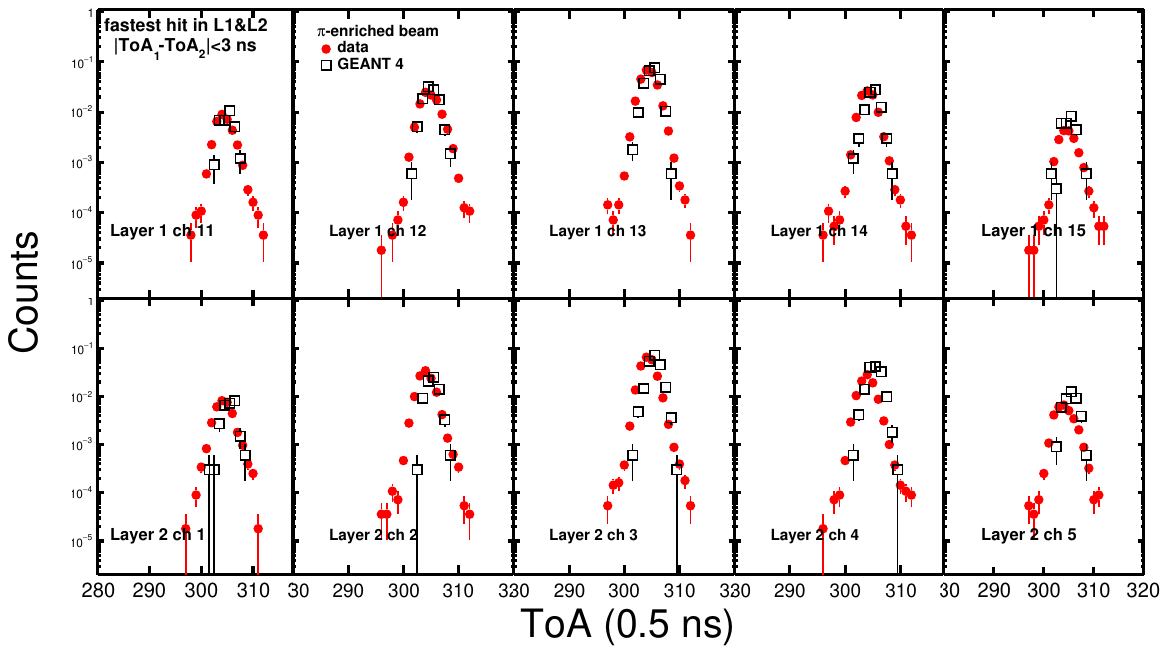}
\caption{Time-of-arrival distributions measured in the 10 channels of the prototype. The upper panel show the results for the first layer, whereas, the bottom panel shows the results for the second layer. The GEANT\,4 results (empty markers) are compared with data (full markers). The measurements were performed using a pion-enriched beam with a momentum of 3\,GeV/$c$. The vertical axis indicates the number of hits normalized to the number of events, which satisfies the selection criteria.}
\label{fig:2b}
\end{figure}

Figures~\ref{fig:2} and~\ref{fig:2b} present the ToT and ToA distributions, respectively, measured across all 10 channels using a 3\,GeV/$c$ pion-enriched beam with a 70\,cm thick absorber. Overall, the MC simulations capture the main aspects of the data very well. The quality of the simulations is the same for both beams as well as for all absorber thicknesses.

For the calculation of the geometrical acceptance using GEANT\,4 simulations, the muon beam was used. The geometrical acceptance for 3\,GeV/$c$ muons was found to be around 87\% for 60\,cm thick absorber and slightly decreases to around 85\% for 100\,cm thick absorber. The dependence is due to scattering that increases with absorber thickness. One has to keep in mind that for the measurement of the hadron suppression, these features are acceptable. However, for the final MID chamber a significantly smaller bar spacing ($<1$\,mm) will be achieved to have better geometrical acceptance than 95\%.

Although only a small fraction of pions is expected to reach the chamber ($\approx1.5$\% assuming the nuclear interaction length equal to 16.77\,cm), a larger contribution could be measured in the experiment due to hadronic showers produced close to the absorber surface or due to kink decays producing secondary muons. Therefore, it is crucial to have a data analysis that can further suppress the background. In order to reduce such a potential background, ML techniques, specifically Boosted Decision Trees (BDT), were used. The analysis is performed using TMVA (toolkit for multivariate data analysis), a toolkit integrated into the ROOT analysis framework that provides a wide range of multivariate (MV) classification algorithms, including BDT. TMVA also offers a comprehensive framework for parallel training, testing, evaluation, and application of MV classifiers, making it particularly suitable for large-scale high-energy physics (HEP) analyses. All MV classifiers have in common to condense (correlated) multi-variable input information into a single scalar output variable, which acts as a powerful discriminator between different classes, such as signal and background~\cite{Hocker:1019880}. The MV input information used for the analysis was based on parameters that are experimentally measured such as hit position ($x$ and $y$), ToT and ToA, in both layers of the MID chamber. The BDT was trained on these experimental variables to discriminate signal (muons) from background (pions).

\begin{figure}[h!]
\centering
\includegraphics[width=0.95\linewidth]{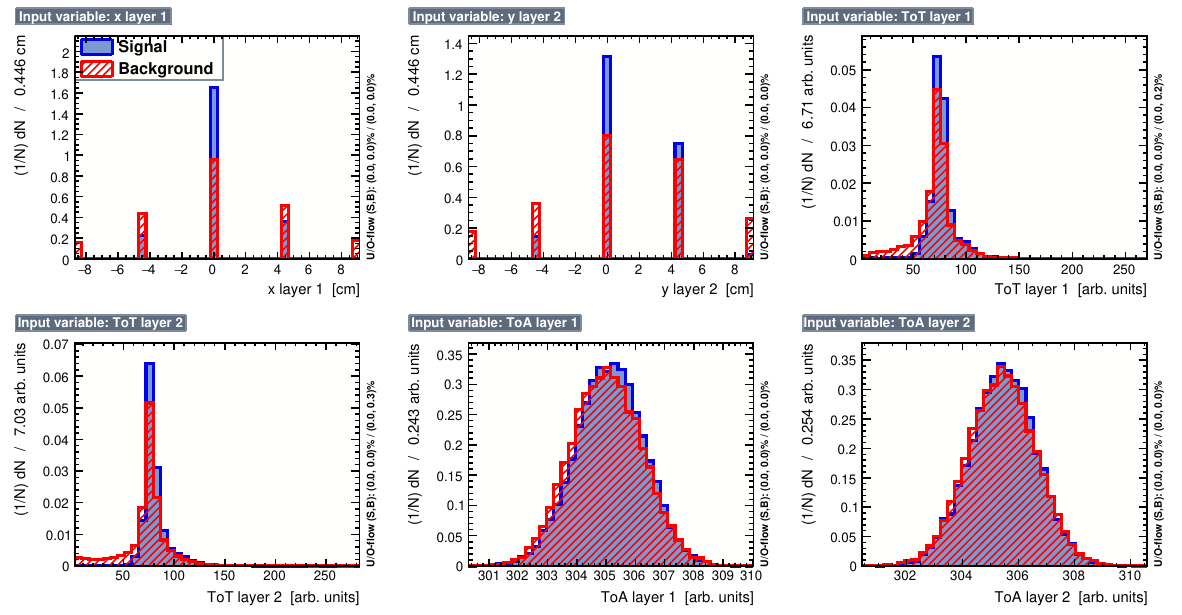}
\caption{The variables used for training the Boosted Decision Trees for the muon tagging in the MID prototype. The position information ($x$ and $y$), ToT and ToA in both two layers are considered. The results were obtained from GEANT\,4 simulations of the test beam setup.}
\label{fig:4}
\end{figure}

For training, 15000 muons and 15000 pions were propagated through the absorber and the detector material. The momentum resolution was assumed to be 1\%, however, a resolution of 20\% was also studied and the results were stable under such a variation. For testing, 8000 pion and 8000 muon guns were used. Figure~\ref{fig:4} shows the distributions associated to signal and background for the training stage. The difference between the two distributions is indicative of the particle showers generated by the pions, as mentioned previously. The signals coming from these showers will not only have a broader spatial distribution when compared to a primary particle but their energy deposition will differ from that of a MIP. This behavior can be confirmed in the signal and background distributions, where one can see a narrower spatial and ToT distribution for the signal case which resembles more the behavior of a primary particle.

The cut on the output BDT was set to the value corresponding to 98\% muon efficiency, this efficiency corresponds to that associated to a single scintillation bar. Once the BDTs were trained they were applied to a statistically independent MC sample aiming at validating the analysis. For this part of the study, a beam consisting of $10^{6}$ particles with a composition of 98\% pions and 2\% muons was simulated.
 
\section{Results
\label{sec:Results}}

\begin{figure}[ht]
\centering
\includegraphics[width=0.31\linewidth]{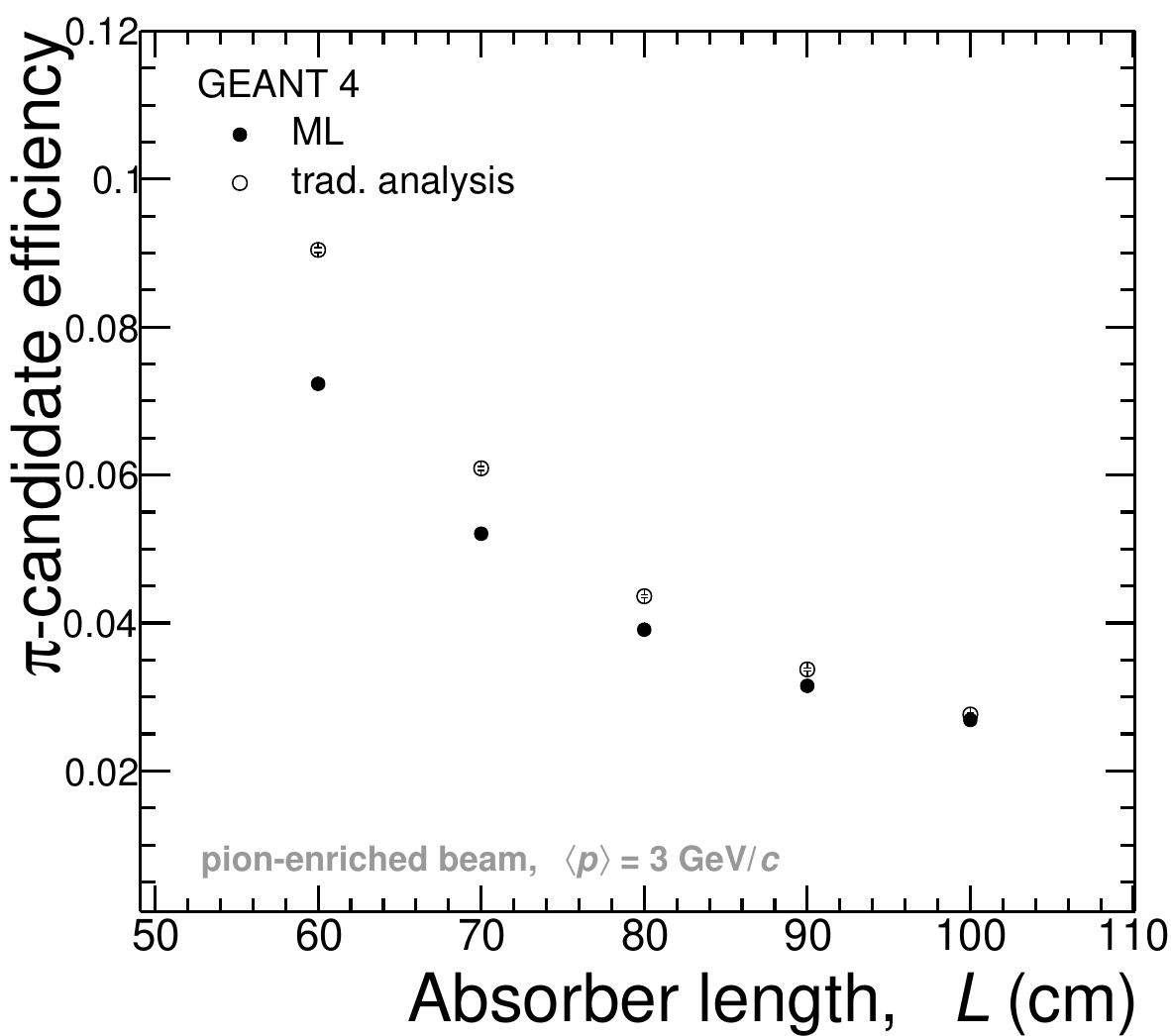}
\includegraphics[width=0.31\linewidth]{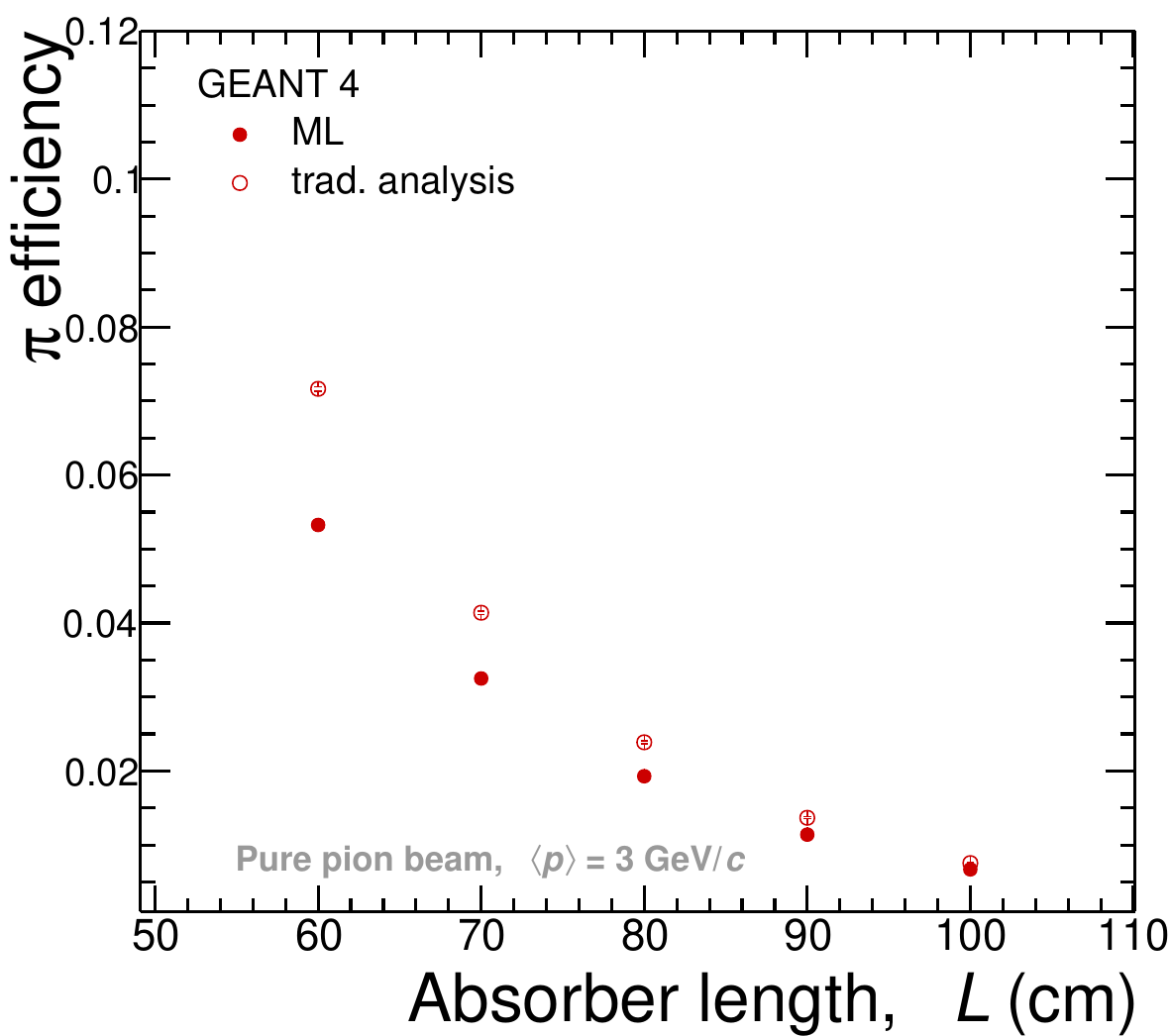}
\includegraphics[width=0.31\linewidth]{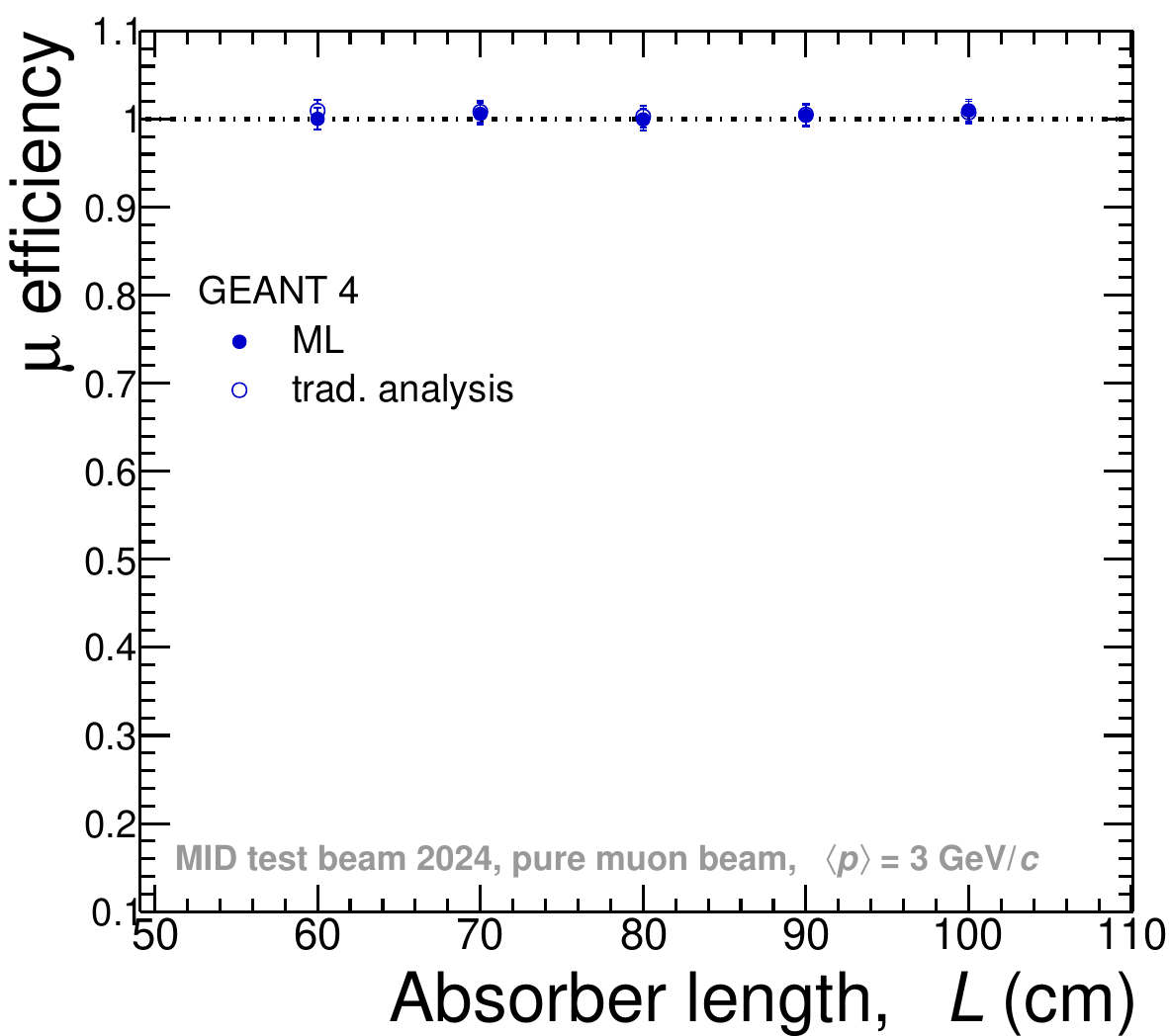}
\caption{(Left) Absorber length dependence of the fraction of muon candidates tagged with BDT in the pion-enriched beam ($p=3$\,GeV/$c$). (Middle) The pion efficiency as a function of absorber length in a pure pion beam. (Right) Similarly the muon efficiency in a pure muon beam as a function of absorber length is presented. Results from traditional analysis based on cuts on ToT (empty markers) are compared with ML results (full markers). The results were obtained using GEANT\,4 simulations.}
\label{fig:5}
\end{figure}

\begin{figure}[ht]
\centering
\includegraphics[width=0.65\linewidth]{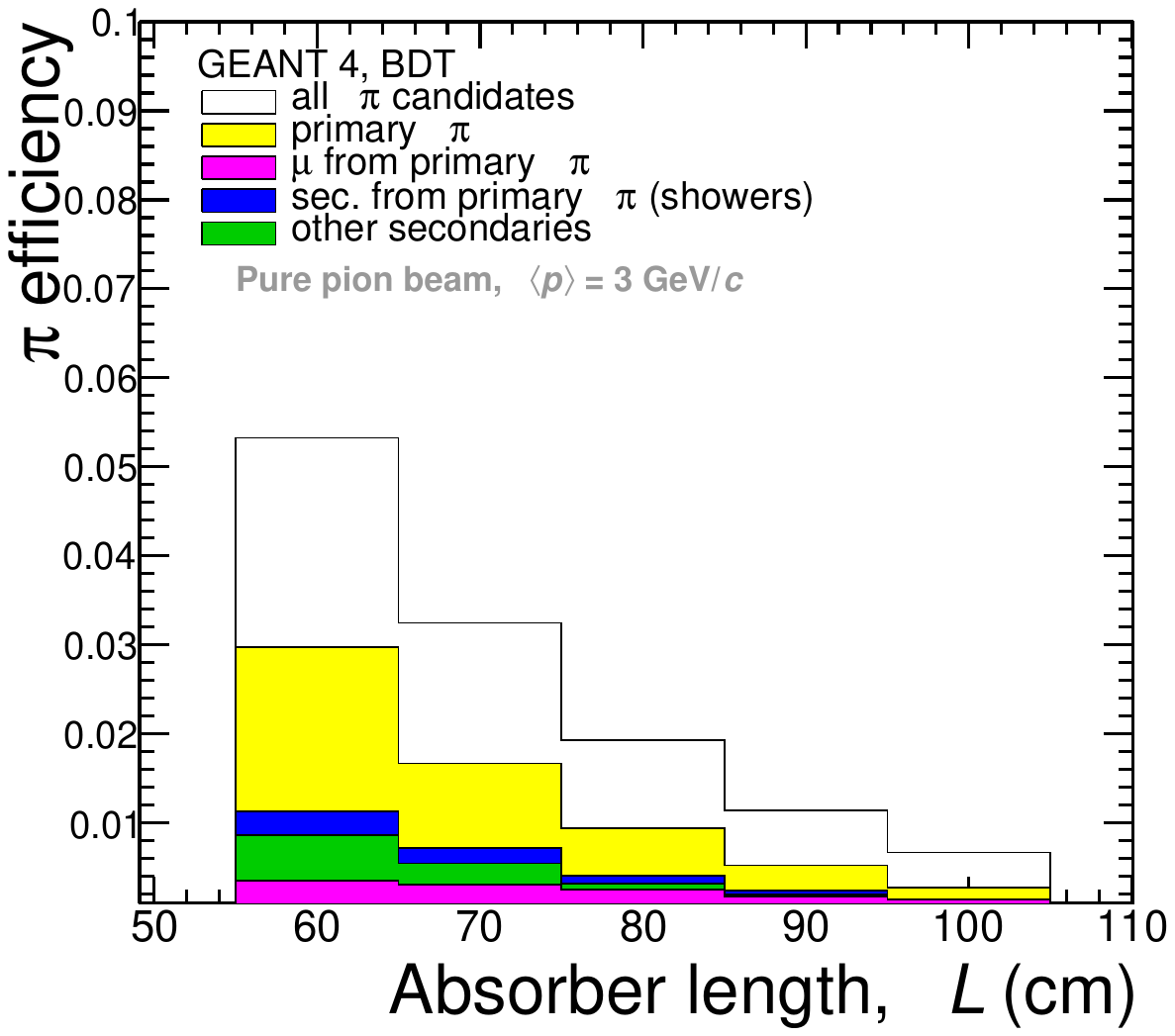}
\caption{Pion-candidate efficiency as a function of absorber thickness using a 3\,GeV/$c$ pure-pion beam. The contributions from primary and secondary particles are displayed.}
\label{fig:5b}
\end{figure}

The pion-candidate efficiency as a function of absorber length in the pion-enriched beam is first evaluated using GEANT\,4 simulation. The efficiency is defined as the fraction of particles that are selected using the trained BDT and thus those that meet the muon selection criteria. For the baseline absorber option, 70\,cm absorber thick, the pion-candidate efficiency amounts to around 5\%. This fraction (and the following ones) is corrected for geometrical acceptance and muon-BDT efficiency. Note that it includes the muon contribution that was injected in the simulated beam. In order to see the discrimination power of BDT, this efficiency is compared to that obtained using a ``traditional'' analysis based on cuts on ToT ($>22.5$\,ns). Figure~\ref{fig:5} shows that ML improves the hadron rejection by about 15\%.  It is worth mentioning that the hadron suppression as a function of absorber length exhibits the characteristic exponential shape. In order to see the real pion efficiency, the pure pion beam was analyzed. The results are presented in the middle panel of Fig.~\ref{fig:5}. For the reference absorber thickness, the pion efficiency amounts to around 3\% using the trained BDT, which is twice  the expected value discussed before (1.5\%). The larger efficiency is attributed to secondary particles produced in showers or muons from kink decays within the absorber. This is illustrated in Fig.~\ref{fig:5b}, which shows the different contributions to the pion-candidate efficiency. For the reference absorber (70\,cm thick), muons contribute with around 15\% and the other secondary particles produced in the absorber contribute with around 34\% to the pion-candidate efficiency. The rest, about 51\%, corresponds to primary pions. Notably, for a thicker absorber (80\,cm thick) the pion efficiency is significantly reduced to 1.9\% with respect to the efficiency achieved with the thickness of the reference absorber. Last but not least, the muon efficiency as a function of absorber length was also evaluated using the muon component of the beam. The plot on the right-hand side of Fig.~\ref{fig:5} shows the efficiencies obtained with both ML and traditional analysis. Both of them were corrected for geometrical acceptance, however, the ML result was further corrected for the BDT efficiency. Both results are very close to each other and fluctuate statistically around 100\%. This result tells us that the detector effects are well under control.

\begin{figure}[ht]
\centering
\includegraphics[width=0.48\linewidth]{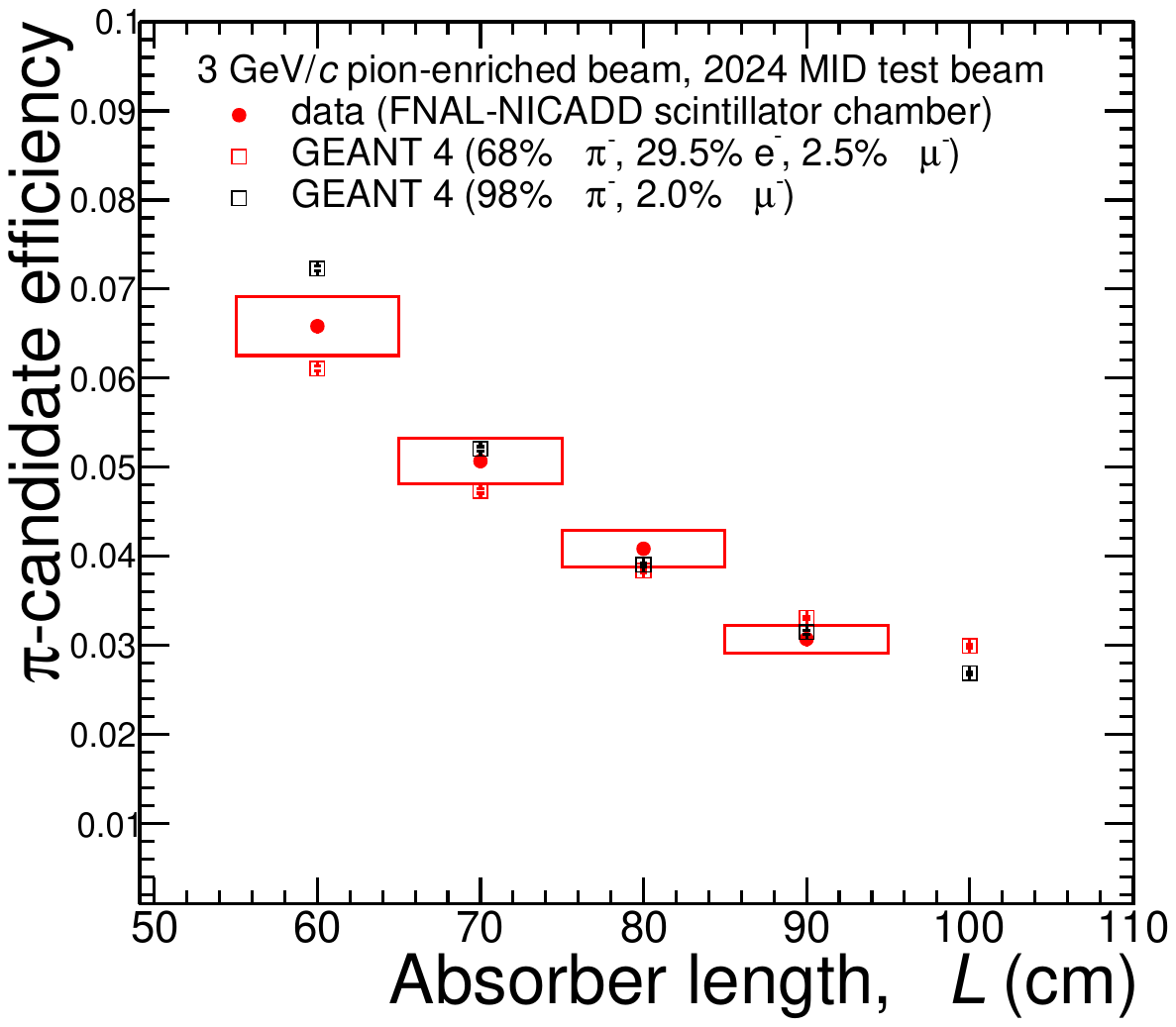}
\includegraphics[width=0.48\linewidth]{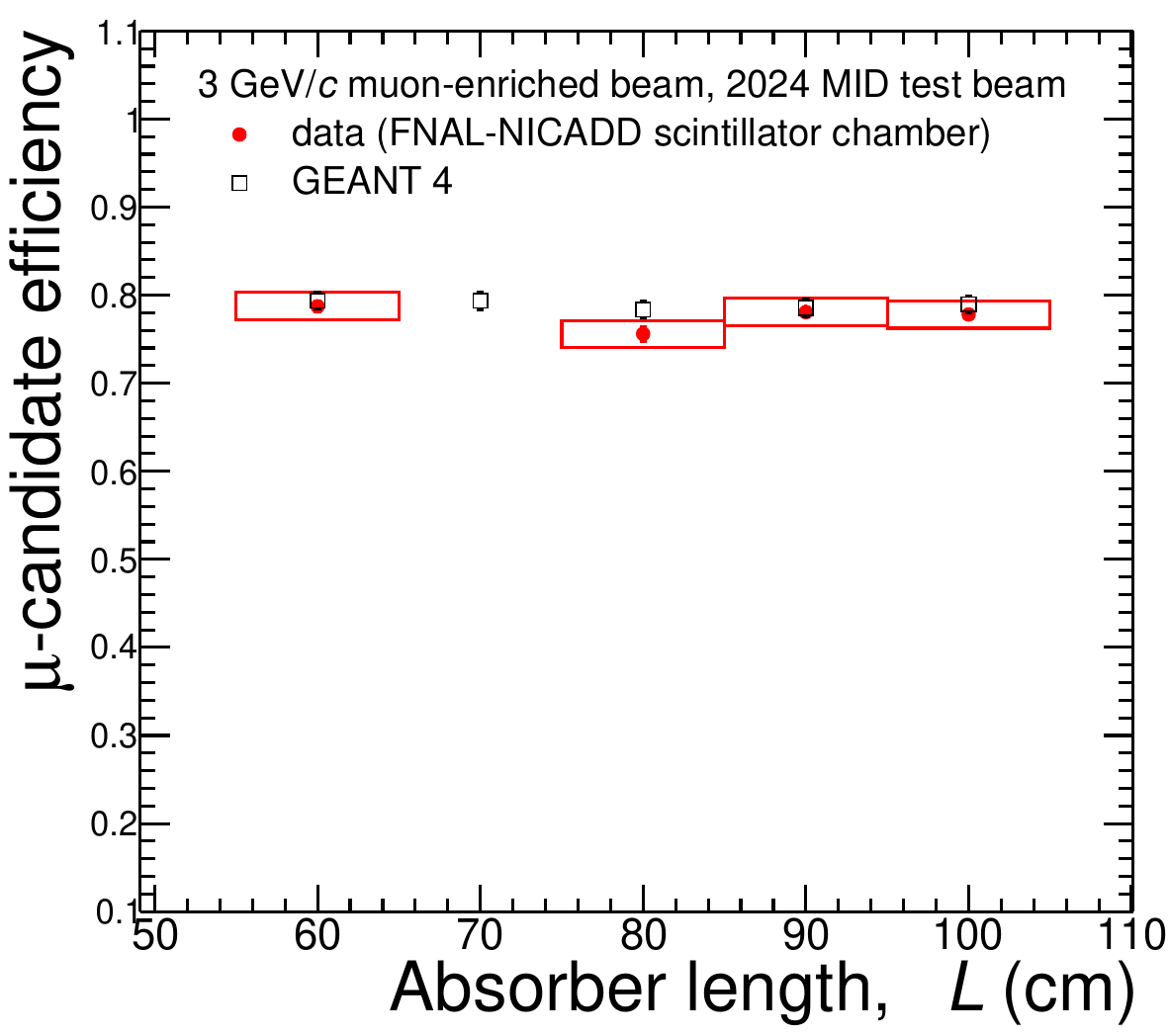}
\caption{Pion-candidate (left) and muon-candidate (right) efficiency as a function of absorber length. Data are compared with MC simulations. Statistical and systematic uncertainties are displayed as error bars and boxes around the data points, respectively. For the pion-enriched beam simulations assuming two beam compositions are shown, whereas a purity of 78\% was assumed in the simulation of the muon-enriched beam.}
\label{fig:6}
\end{figure}

The measured pion-candidate efficiency as a function of absorber length and applying the trained BDTs to the pion-enriched beam is shown in Fig.~\ref{fig:6}. The systematic uncertainty was assigned based on the difference between the efficiency measured with the MID chamber and that obtained using the array of MWPCs. The independent analysis (MWPC data) used position information measured in up to 3 MWPCs placed behind the absorber. The information is used to define a $\chi^2$ function that is optimized to maximize the muon detection efficiency. A cut is applied at the optimized $\chi^2$ value, and then the pion-candidate efficiency is evaluated. The difference between the two analyses is around 5\%, which is assigned as systematic uncertainty.  Data are compared with the pion-enriched beam simulated with GEANT\,4. Originally, a beam composition consisting of 98\% pions and 2\% muons was assumed in the simulation, and the model was found to describe the data. However, a recent technical document suggests that in addition to the muon contamination, an electron contribution from 25\% up to 30\% can be present in the negative-charged pion beam~\cite{vanDijk:2025ggb}. A pion-enriched beam consisting of 68\% pions, 29.5\% electrons and 2.5\% muons was simulated, and our data is found to be consistent with this assumption within one sigma. Based on the discussion above, and given the agreement between the data and Monte Carlo simulations, the measured pion efficiency is around $3.0\pm0.15$\% (70\,cm absorber thickness) that is consistent with the requirements for the MID detector and gives a competitive $J/\psi$ signal-to-background and significance both for pp and Pb--Pb collisions at the expected interaction rates in ALICE\,3~\cite{alice3loi,Dainese:2925455}. For completeness, the muon-candidate efficiency as a function of absorber length is also shown. MC simulations assuming a muon purity of 78\% describe the data. Our result suggests that the muon-enriched beam delivered at the PS T10 has a purity of $78.0\pm1.6$\% for momentum of 3\,GeV/$c$. The systematic uncertainty of 2\% is assigned based on the difference with respect to the independent analysis using the array of MWPCs discussed above.  

\section{Summary and outlook \label{sec:Conclusions}}

A test of a scintillator-based muon chamber for the MID subsystem of ALICE\,3 was conducted at the CERN T10 test beam facilities. The prototype consist of FNAL-NICADD scintillator bars equipped with Kuraray wavelength shifting fiber and readout with Hamamatsu SiPM. For the small-size ALICE\,3 MID prototype, the length of the bar was 25\,cm and the bar spacing was 3.5\,mm. Pion- and muon-enriched beams were used. The experimental setup was simulated with GEANT\,4, the simulations were used to train ML algorithms that were further applied to the data. The goal of the test was the measurement of the hadron suppression considering different iron absorber lengths. For a muon efficiency of 98\% in a single bar, a pion-suppression factor of $\approx30\pm1.5$ is found for 70\,cm-thick absorber. The pion-suppression factor can be increased to $\approx50\pm2.5$ and $\approx100\pm5$ for 80\,cm- and 90\,cm-thick absorber, respectively. Regarding the features of the beams, GEANT\,4 simulations suggest that the purity of the muon-enriched is roughly $78.0\pm1.6$\%. Whereas, for the pion enriched beam our data support the presence of a fraction of muons of up to 2.5\%, and electron contamination of up to 30\%. 

In the near future, an improved prototype will be built and tested. The aim is to achieve a geometrical acceptance higher than 95\% by reducing dead areas. Moreover, the size of the chamber will be $1\times1$\,m$^{2}$. Other variables such as hit multiplicity will be explored to improve muon tagging and further reduce the background.


\appendix


\section*{Acknowledgments}
The authors acknowledge the technical support from Miguel Enrique Pati\~no Salazar, Luciano D\'iaz Gonz\'alez, Jes\'us Eduardo Murrieta Le\'on, Sa\'ul Aguilar Salazar, and Jaime Everardo P\'erez Rodr\'guez. The participation of Irandheny~Yoval~Pozos, Victor V\'azquez Campos, Timea~Szollosova, Matej Tonka, and Rafael Narcio during the shifts or prototype construction is acknowledge. The MID-chamber construction and data processing were supported through funding from PAPIIT-UNAM (Grant IG100524) and PAPIME-UNAM (Grant PE100124). The construction of the MID prototype, their corresponding simulations and data analysis were done at UNAM. The construction of MWPC was supported by the Hungarian NRDI Fund research grant TKP2021-NKTA-10 and NKFIH OTKA FK131979. MWPC construction and testing was completed within the Vesztergombi Laboratory for High Energy Physics (VLAB) at HUN-REN Wigner RCP. S.~R. acknowledges the support from the Grant Agency of the Czech Technical University in Prague, Grant Number SGS24/145/OHK4/3T/14. E.~G. acknowledges the support from the Chicago State University, NSF award PHY-2208883.


\bibliography{ref}

\end{document}